\documentclass[preprint,prb,showpacs]{revtex4}
\usepackage{amsmath}    % need for subequations 
\usepackage{graphicx}   % need for figures 
\usepackage{verbatim}   % useful for program listings 
\usepackage{color}      % use if color is used in text 
\usepackage{subfigure}  % use for side-by-side figures 
\usepackage[pdffitwindow,colorlinks,citecolor={red},linkcolor={blue}]{hyperref}
\usepackage[dvipsnames]{xcolor}
\usepackage{multirow}
\usepackage{soul}
\def\tama{10cm}
\begin{document}

\title{Pure Spin Current Injection in Hydrogenated Graphene Structures}
\author{Reinaldo Zapata-Pe\~na}
\affiliation{Centro de Investigaciones en \'Optica, Le\'on,
Guanajuato 37150, M\'exico}
\author{Bernardo S. Mendoza}
\email[E-mail: ]{bms@cio.mx}
\affiliation{Centro de Investigaciones en \'Optica, Le\'on,
Guanajuato 37150, M\'exico}
\author{Anatoli I. Shkrebtii}
\affiliation{University of Ontario, Institute of Technology,
Oshawa, ON, L1H 7L7, Canada}

\date{\today}

\begin{abstract}
We present a theoretical study of spin-velocity injection (SVI) of a pure spin 
current (PSC) induced by  a linearly polarized light that impinges normally on 
the surface of  two 50\% hydrogenated noncentrosymmetric two-dimensional (2D) 
graphene structures. The first structure, hydrogenated at only one side, 
labeled Up, also known as graphone, and the second, labelled Alt, is 
25\% hydrogenated at both sides. The hydrogenation opens an energy gap in both 
structures. We analyze two possibilities: in the first, the spin is fixed 
along a chosen direction, and the resulting SVI is calculated; 
in the second, we choose the SVI direction along 
the surface plane, and calculate the resulting 
spin orientation. This is done by changing the energy 
$\hbar\omega$ and polarization angle $\alpha$  of the incoming  light. The 
results are calculated within a full electronic band structure scheme using the 
Density Functional Theory (DFT) in the Local Density Approximation (LDA). The 
maxima of the spin-velocities are reached when $\hbar\omega=0.084$\,eV and 
$\alpha=35^\circ$ for the Up structure, and $\hbar\omega=0.720$\,eV and 
$\alpha=150^\circ$ for the Alt geometry. We find a speed of 668\,Km/s and 
645\,Km/s for the Up and the Alt structures, respectively, when the spin points 
perpendicularly to the surface. Also, the response is maximized by 
fixing the spin-velocity direction 
along a high symmetry  axis, obtaining a speed of 688Km/s 
with the spin pointing at $13^\circ$ from the surface normal, for the Up, 
and 906 Km/s and the spin pointing at $60^\circ$ from the surface normal, for 
the Alt system. These speed values are of order of magnitude larger than those of bulk semiconductors, 
such as CdSe and GaAs, thus making the hydrogenated graphene structures 
excellent candidates for spintronics applications. 
\end{abstract}


\maketitle

\section{Introduction}
\label{sec:introduction}

Spintronics is an emerging research field of electronics in which the
manipulation and transport of the electron spin in a solid state materials  is
central, adding a new degree of freedom to conventional charge
manipulation.\cite{wolfSC04,fabianAPS07} At present, there is an increasing
interest in attaining the same level of control over the transport of spin at
micro- or nano-scales, as it has been done for the flow of charge in typical
3D-bulk based electronic devices.\cite{awschalomNP2007} Several semiconductor
spintronics devices have been proposed \cite{majumdarAPL06,
dattaAPL90,gotteNat16,pershinPRB08}, and some of them require spin polarized
electrical current \cite{awschalomSSBM13} or pure spin current (PSC). One of
the difficulties in creating measurable  spin current and
development of PSC based semiconductor devices is the fact that the spin
relaxation time in conventional semiconducting materials cloud be too short to enable the
spin transport, and may result in a non-observable spin
current.\cite{murakamiSc03} For PSC there is no net motion of charge; spin-up
electrons move in a given direction, while spin-down electrons travel in the
opposite one. This effect can be due to one-photon absorption of linearly
polarized light by a semiconductor, with filled valence bands and empty
conduction bands, illuminated by light with photon energy larger than the
energy gap. This phenomenon can be due to spin injection,\cite{malPRB03} Hall
Effects,\cite{sinovaPRB04} interference of two optical beams,\cite{bhatPRL00,
najmaiePRB03} or one photon absorption of linearly polarized
light\cite{bhatPRL05}. The last effect has been observed in gallium arsenide
(GaAs),\cite{zhaoPRL2006, stevensPRL03} aluminum-gallium arsenide
(AlGaAs),\cite{stevensPRL03} and Co$_2$FeSi.\cite{kimuraNGPAM12}

The spin velocity injection (SVI) is an optical effect that quantifies the velocity at which
a PSC moves along the direction $\hat{\mathbf{a}}$, 
with the spin of the electron polarized along the direction $\hat{\mathbf{b}}$.
One photon absorption of polarized
light produces an even distribution of electrons in $\mathbf{k}$
space, regardless of the symmetry of the material, resulting in a null
electrical current.\cite{bhatPRL05} 
Then, the electrons excited to the
conduction bands at opposite $\mathbf{k}$ points will result in opposite spin
polarizations producing no net spin injection in centrosymmetric 
materials.\cite{bhatPRL05} 
If the crystalline structure of the material
is noncentrosymmetric, the spin polarization injected at a given $\mathbf{k}$
point not necessarily vanishes.\cite{alvaradoPRL85,schmiedeskampPRL88}
Therefore, since the velocities of electrons at opposite $\mathbf{k}$ points
are opposite, a PSC will be produced. 

Graphene, an allotrope of carbon with hexagonal 2D lattice structure,
demonstrates properties such as fractional quantum Hall effect at room
temperature, excellent thermal transport properties, excellent
conductivity\cite{heerscheNat07} and strength \cite{geimNM07, reinaNL08,
novoselov2S07, balandinNL08}, being a perfect platform  for two-dimensional (2D)
electronic systems; however, numerous important electronic applications are disabled by the
absence of a semiconducting gap. Recent studies demonstrate that a narrow band
gap  can be opened in graphene by applying an electric field,\cite{zhangN09}
reducing the surface area,\cite{hanPRL07} or applying uniaxial
strain.\cite{niACSN08} Another possibility to open the gap is by doping; this
has been successfully achieved using nitrogen,\cite{weiNL2009} 
boron-nitrogen,\cite{guoIJ11} silicon,\cite{colettiPRB10} 
noble-metals,\cite{varykhalovPRB10} and hydrogen.\cite{eliasS09, guisingerNL09,
samarakoonACSN10} Depending on the percentage of hydrogenation and spatial
arrangements of the hydrogen-carbon bonds, hydrogenated graphene demonstrates
different structural configurations and a tunable electron gap, as it has been
proven in Ref. \onlinecite{shkrebtiiPSSC12}. 

\begin{figure}[ht!]
    \centering
    \includegraphics[width=5cm]{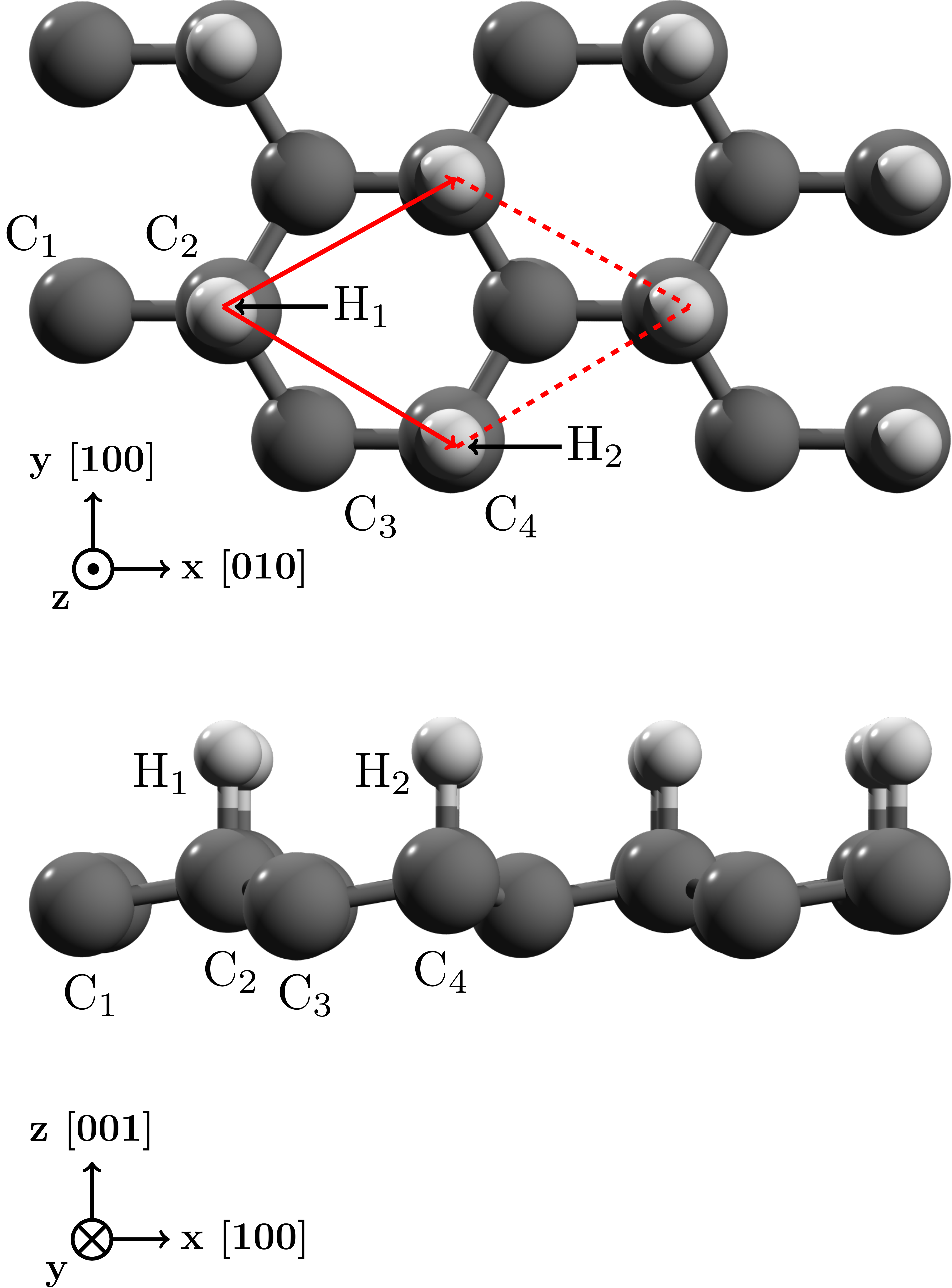}
    \caption{(color online) Top (top panel) and side (bottom panel) views of
the Up structure along with the Cartesian $x$, $y$, and $z$ directions. The
dark (light) spheres are the C (H) atoms. The primitive hexagonal unit cell is
also shown.}
    \label{fig:up-struc}
\end{figure}

\begin{figure}[ht!]
    \centering
    \includegraphics[width=5cm]{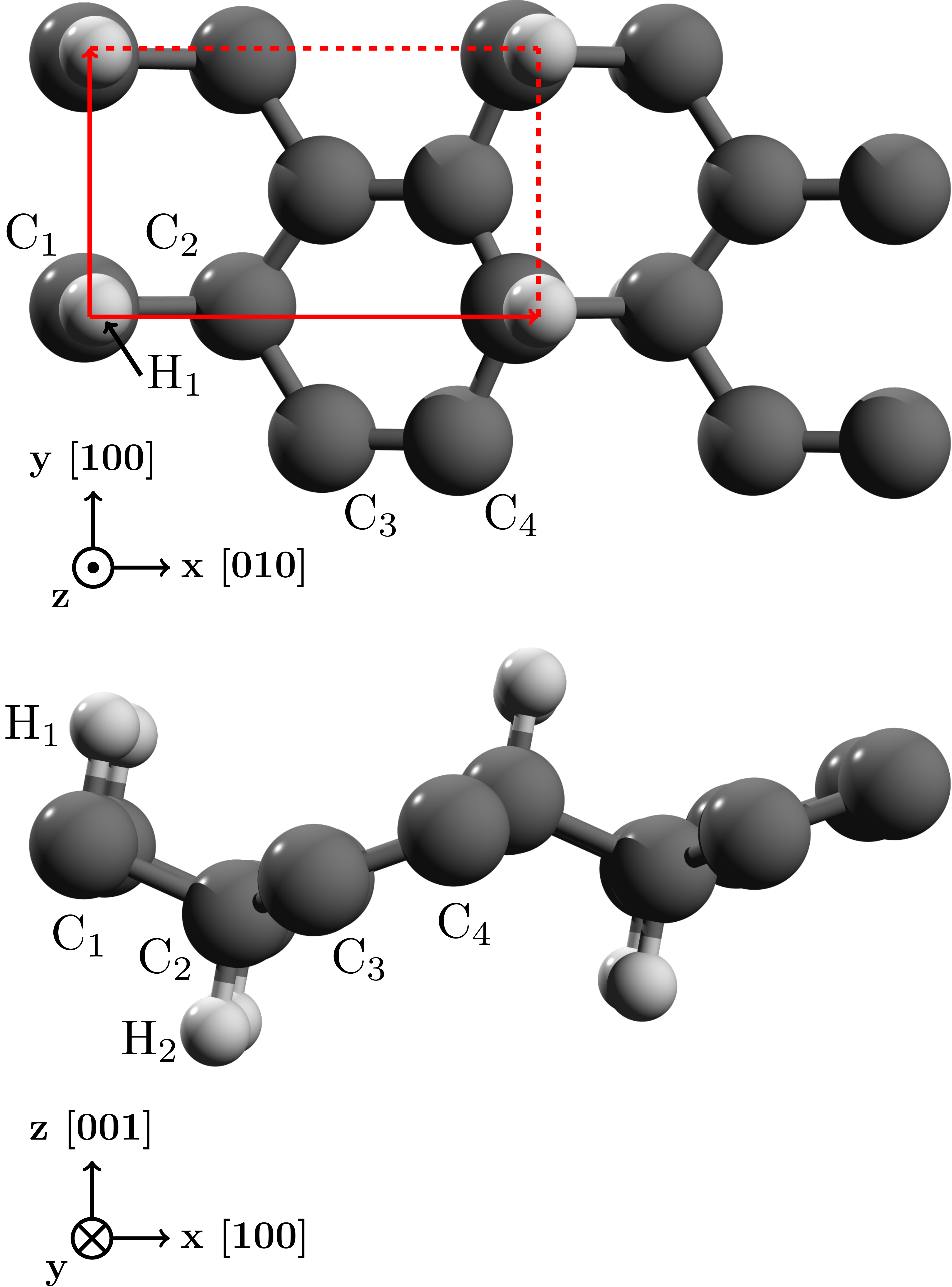}
    \caption{(color online) Top (top panel) and side (bottom panel) views of
the Alt structure along with the Cartesian $x$, $y$, and $-z$ directions. The
dark (light) spheres are the C (H) atoms. The primitive rectangular unit cell
is also shown. }
    \label{fig:alt-struc}
\end{figure}

In this paper, we offer two 50\% hydrogenated graphene
noncentrosymmetric structures, both demonstrating a discernible band gap. The
first one, labelled as the Up structure, also known as
graphone,\cite{gmitraPRL13}  has
hydrogen atoms bonded to the carbon layer only on the upper side of the
structure; we consider here the magnetic isomer of graphone, with the so-called
``chair'' structure  shown in Fig. \ref{fig:up-struc}.
In contrast,  the Alt structure, shown in Fig. 
\ref{fig:alt-struc}, has hydrogen alternating on the upper and bottom sides of 
the carbon sheet.\cite{zapataPSB2016}

Both the Up and the Alt  structures
are noncentrosymmetric, and therefore,
they are good candidates in which SVI can be
induced. In this article, 
we address theoretically the spin-velocity injection
by one-photon absorption of linearly polarized light, analyzing
in our structures
two possible scenarios of practical interest. 
The first case is by fixing the spin of the electrons along $z$, i.e.,
perpendicular 
to the surface plane, with the resulting velocity
directed along the surface of the
structures on the $xy$ plane. 
In the second case we fix the SVI velocity along the
$x$ or $y$ direction, and then, the resulting spin is directed outward of the
$xy$ plane.


This paper is organized as follows. In Section \ref{sec:theory} we outline the
formalism and the main expressions  that describe PSC and SVI. In Section
\ref{sec:results} we describe the numerical details  and  discuss the
corresponding SVI spectra for the Up and Alt structures. Finally, we summarize
our findings in Section \ref{sec:conclusions}.

\section{Theory}
\label{sec:theory}

In this section, we  summarize the theoretical approach, involved in the
calculation of the spin velocity injection (SVI) resulting from the pure spin
current (PSC).
 
To calculate the velocity of the spin injection
${\cal V}^{\mathrm{a}\mathrm{b}}(\omega)$ along the direction $\hat{\mathbf{a}}$, at
which the spin moves in a
polarized state along direction $\hat{\mathbf{b}}$,
we start with the operator that describes the electronic SVI, written as
\begin{align}\label{z.1}
\hat K^{\mathrm{a}\mathrm{b}} = 
\frac{1}{2}\left( \hat v^\mathrm{a} \hat S^\mathrm{b} 
+\hat  S^\mathrm{b} \hat v^\mathrm{a}\right) 
.
\end{align} 
Here $\hat{\mathbf v}=[\hat{\mathbf r},\hat H_0]/i\hbar$ 
is the velocity operator, with
$\hat {\mathbf r}$ 
being the position operator and $\hat H_0$ the unperturbed
ground state Hamiltonian; 
the Roman superscripts  indicate Cartesian coordinates. 
To obtain the expectation value of 
$\hat K^{\mathrm{a}\mathrm{b}}$, we use the length gauge for the perturbing
Hamiltonian, written as
\begin{align}\label{z.2}
\hat H_{\text{p}}=-e\hat{\mathbf r}\cdot{\mathbf E}(t)
,
\end{align}   
where the applied electric field of the beam of light is given by
\begin{align}\label{z.3}
{\mathbf E}(t) = 
{\mathbf E}(\omega)e^{-i\omega t} + {\mathbf E}^*(\omega)e^{i\omega t}
.
\end{align}
In order to calculate the response of the system to ${\mathbf E}(t)$, one needs
to take into account the excited coherent superposition of the spin-split
conduction bands inherent to the noncentrosymmetric semiconductors considered
in this work. 
To include the coherence, we follow Ref.~\onlinecite{nastosPRB07}
and use a multiple scale approach that solves the equation of motion for the
single particle density matrix 
$\hat{\rho}({\mathbf k};t)$,  
leading to
\begin{align}\label{z.4}
&\frac{\partial \rho_{cc'}({\mathbf k;t})}{\partial t} =
\frac{e^{2}E^{\mathrm{a}}(\omega)E^{\mathrm{b*}}(\omega)}
{i \hbar^{2}}
\sum_{v}r^{\mathrm{a}}_{cv}({\mathbf k}) r^{\mathrm{b}}_{vc'}({\mathbf k})
\left( \frac{1}{\omega - \omega_{c'v}({\mathbf k}) - i \epsilon} 
- 
\frac{1}{\omega - \omega_{cv}({\mathbf k}) + i \epsilon} \right)
,
\end{align}
where we assumed that the conduction bands $c$ and $c'$ are quasi-degenerate
states, and we take $\epsilon\to 0$ at the end of the calculation. 
Since the 
spin-splitting of the valence ($v$) bands is very small, we neglect it throughout
this work,\cite{nastosPRB07} and then
$\rho_{vv'}({\mathbf k};t)= 
\rho_{cv}({\mathbf k};t) =0$. 
 The matrix elements of any operator ${\cal O}$ are
given by ${\cal O}_{nm}({\mathbf k})=\langle n{\mathbf k}|\hat{\cal O}|
m{\mathbf k}\rangle$, where $H_{0}|n{\mathbf k}\rangle = \hbar
\omega_{n}({\mathbf k})|n{\mathbf k}\rangle$ with $\hbar \omega_{n}({\mathbf
k})$ being the energy of the electronic band $n$ and $m$ at point ${\mathbf k}$ in the
irreducible Brillouin zone (IBZ),  $|n{\mathbf k}\rangle$ is the Bloch state,
and $\omega_{nm}({\mathbf k})=\omega_{n}({\mathbf k})-\omega_{m}({\mathbf k})$.
Using $\mathcal{O} = \mathrm{Tr}(\hat{\rho}\hat{\mathcal{O}})$ for the
expectation value of an observable $\mathcal{O}$, where $\mathrm{Tr}$ denotes
the trace, we obtain
\begin{align}\label{z.5}
\mathcal{O} = & 
\int \frac{d^{3}k}{8\pi^{3}} \sum_{cc'} \rho_{cc'}({\mathbf k}) 
\mathcal{O}_{c'c}({\mathbf k}),
\end{align}
where we used the closure relationship $\sum_{n}|n{\mathbf k}\rangle \langle
n{\mathbf k}| = 1$, where $n$ goes over all $v$ and $c$ states.
Therefore, using  Eqs. \eqref{z.4}
and \eqref{z.5}, the rate of change of $\mathcal{O}$, $\dot{\mathcal{O}} =
\mathrm{Tr} \left( \dot{\hat\rho} \hat{\cal O} \right)$, is given by
\begin{align}
\dot{\mathcal{O}} 
&=\frac{e^{2}}{i\hbar^{2}} \int \frac{d^{3}k}{8\pi^{3}} 
\sum'_{cc'} \mathcal{O}_{c'c}({\mathbf k}) 
r^{\mathrm{a}}_{cv}({\mathbf k})  r^{\mathrm{b}}_{vc'}({\mathbf k})  
\left( \frac{1}{\omega - \omega_{c'v}({\mathbf k})  - i\epsilon} - 
\frac{1}{\omega - \omega_{cv}({\mathbf k})  + i\epsilon} \right)
E^{\mathrm{a}}(\omega) E^{\mathrm{b*}}(\omega)
\label{eq:dotO}
.
\end{align}
The prime symbol $'$ in the sum means that $c$ and $c'$ are quasi-degenerate
states, and the sum only covers these states. Replacing  $\hat{\cal O}
\rightarrow \hat{K}^{\mathrm{ab}}$, in the above expression, one can show that
\begin{equation}
\dot{K}^{\mathrm{ab}}(\omega) =
\mu^{\mathrm{abcd}}(\omega)
E^{\mathrm{c}}(\omega) E^{\mathrm{d*}}(\omega),
\label{eq:dotk}
\end{equation}
where the repeated Cartesians upperscripts  are summed, and 
\begin{equation}\label{eq:mu}
\begin{aligned}
\mu^{\mathrm{abcd}}  (\omega) 
=
\frac{\pi e^{2}}{\hbar^{2}} \int 
\frac{d^{3}k}{8 \pi^{3}} \sum'_{vcc'}
\delta(\omega-\omega_{cv}({\mathbf k}) 
\mathrm{Re} \left[ K^{\mathrm{ab}}_{cc'}({\mathbf k}) 
\left(  
r^{\mathrm{c}}_{vc'}({\mathbf k})   
r^{\mathrm{d}}_{cv }({\mathbf k})  +
(c \leftrightarrow d)  
\right) 
\right]
\end{aligned}
\end{equation} 
is the pseudotensor that describes the rate of change of the PSC in
semiconductors. To derive what we presented above we used $
K^{\mathrm{ab}}_{nm}(-\mathbf{k}) = K^{\mathrm{ab*}}_{nm}({\mathbf k})$, which
follows from time-reversal invariance. Since $\mu^{\mathrm{abcd}}(\omega)$ is
real, we have that $\mu^{\mathrm{abcd}}(\omega) =
\mu^{\mathrm{abdc}} (\omega)$. We point out that Eq.~\eqref{eq:mu} is
identical to Eq. (3) of Ref. \onlinecite{bhatPRL05} derived using the
semiconductor optical Bloch equations. Using the closure relation,
\begin{equation}
K^{\mathrm{ab}}_{cc'}({\mathbf k}) = \frac{1}{2}
\sum_{l=v,c}
\left(v^{\mathrm{a}}_{cl}({\mathbf k})S^{\mathrm{b}}_{lc'}({\mathbf k})
+S^{\mathrm{b}}_{cl}({\mathbf k}) v^{\mathrm{a}}_{lc'}({\mathbf k})
\right)
.
\label{eq:velspimatelem}
\end{equation}

We define the spin velocity injection (SVI) as
\begin{equation}\label{eq:vab-w}
\mathcal{V}^{\mathrm{ab}}(\omega) \equiv
\frac{\dot{K}^{\mathrm{ab}}(\omega)}{(\hbar/2) \dot{n}(\omega)},
\end{equation}  
which gives the velocity, along direction $\hat{\mathbf{a}}$, at which the spin moves in a
polarized state along direction $\hat{\mathbf{b}}$. 
The carrier injection rate $\dot n(\omega)$ is written as\cite{nastosPRB07}
\begin{equation}
\dot{n}(\omega) =
\xi^{\mathrm{ab}}(\omega) E^{c }(\omega) E^{d*}(\omega)
\label{eq:dotn}
\end{equation}
where the tensor 
\begin{equation}\label{eq:xi}
\begin{aligned}
\xi^{\mathrm{ab}}(\omega)
&
=
\frac{2\pi e^{2}}{\hbar^{2}} \int 
\frac{d^{3}k}{8 \pi^{3}}
 \sum_{vc}
r^{\mathrm{a}}_{vc'}({\mathbf k})  
r^{\mathrm{b}}_{cv }({\mathbf k})  
\delta(\omega-\omega_{cv}({\mathbf k})), 
\end{aligned}
\end{equation}
is related to the imaginary part of the linear optical response tensor by
$\mathrm{Im} [\epsilon^{\mathrm{a}\mathrm{b}}(\omega)] =
2\pi\epsilon_0\hbar\xi^{\mathrm{a}\mathrm{b}}(\omega)$.

The function ${\cal V}^{\mathrm{a}\mathrm{b}}(\omega)$ allows us to quantify
two very important aspects of PSC. On one hand, we can fix the spin direction
along $\hat{\mathbf{b}}$ and calculate the resulting electron velocity. On the other
hand, we can fix the velocity of the electron along $\hat{\mathbf{a}}$
 and study the
resulting direction along which the spin is polarized. To this end, the
additional advantage of  2D structures, besides being noncentrosymmetric, is
that we can use an incoming linearly polarized  light at normal incidence, and
use the  direction of the polarized  electric field to control ${\cal
V}^{\mathrm{a}\mathrm{b}}(\omega)$. Indeed, writing ${\mathbf E}(\omega) =
E_0(\omega)(\cos\alpha\,\hat{\mathbf x}+\sin\alpha\,\hat{\mathbf y})$, where
$\alpha$ is the polarization angle, we obtain from Eq. \eqref{eq:vab-w} that
\begin{widetext}
\begin{align}
\mathcal{V}^{\mathrm{ab}}(\omega,\alpha)
&= 
\frac{2}{\hbar\xi(\omega)}
\left(\mu^{\mathrm{abxx}}(\omega)\cos^{2}\alpha + 
\mu^{\mathrm{abyy}}(\omega)\sin^{2}\alpha + 
\mu^{\mathrm{abxy}}(\omega)\sin 2\alpha\right)
,
\label{eq:vab-aw}
\end{align}
\end{widetext}
since for the structures chosen in this article,
$\xi^{\mathrm{xx}}(\omega)=\xi^{\mathrm{yy}}(\omega)\equiv\xi(\omega)$, and
$\xi^{\mathrm{xy}}(\omega)=0$. Next, we identify two options for ${\cal
V}^{\mathrm{a}\mathrm{b}}(\omega)$.

\subsection{Fixing the spin polarization}\label{sec:theory-fixspin}

Analyzing the SVI, Eq. \eqref{eq:vab-aw}, we calculate the magnitude
of the electron velocity along the plane of the structure, with the spin
polarized along $\hat{\mathbf{b}}$ direction as
\begin{equation}
\mathcal{V}_{\sigma^{\mathrm{b}}}(\omega,\alpha)
\equiv
\sqrt{
\left(\mathcal{V}^{\mathrm{xb}}(\omega,\alpha)\right)^{2}\ +
\left(\mathcal{V}^{\mathrm{yb}}(\omega,\alpha)\right)^{2}\ 
}, 
\label{eq:vs-mag}
\end{equation}
and define the angle at which the velocity is directed on the $xy$ plane as
\begin{equation}
\gamma_{\sigma^\mathrm{b}} (\omega,\alpha)
=
\tan^{-1} \left( \frac{\mathcal{V}^{\mathrm{yb}}(\omega,\alpha)}
{\mathcal{V}^{\mathrm{xb}}(\omega,\alpha)} \right)
.
\label{eq:gamma-ang}
\end{equation}
We also define two special angles
\begin{equation}
\gamma_{\sigma^\mathrm{b}}^\parallel(\omega,\alpha) = \alpha, 
\label{eq:gamma-par} 
\end{equation}
and
\begin{equation}
\gamma_{\sigma^\mathrm{b}}^\perp(\omega,\alpha) = \alpha \pm 90^{\circ},
\label{eq:gamma-perp}
\end{equation}
corresponding to the electron velocity being parallel or perpendicular to the
incoming light polarization direction, respectively. 
The subscript $\sigma^\mathrm{b}$
denotes the spin along $\hat{\mathbf{b}}$.

\subsection{Fixing the electron velocity.}\label{sec:theory-fixvel}

Fixing the calculated velocity along $\mathrm{a}=x$ or $\mathrm{a}=y$, we
define its corresponding magnitude as
\begin{align}
\mathcal{V}_{\mathrm{a}}(\omega,\alpha) \equiv 
\sqrt { 
\left(\mathcal{V}^{\mathrm{ax}}(\omega,\alpha)\right)^{2} +
\left(\mathcal{V}^{\mathrm{ay}}(\omega,\alpha)\right)^{2} +
\left(\mathcal{V}^{\mathrm{az}}(\omega,\alpha)\right)^{2} 
},
\label{eq:vv-mag}
\end{align}
from where we see that the spin would be oriented in the $xyz$ system of
coordinates along the polar angle, 
\begin{equation}
\theta_{\mathrm{a}}  (\omega,\alpha) = 
\cos^{-1} \left( \frac{\mathcal{V}^{\mathrm{az}}(\omega,\alpha)}
{\mathcal{V}_{\mathrm{a}}(\omega,\alpha)} \right),
 \quad 0 \leq \theta \leq \pi, 
\label{eq:polar-ang}
\end{equation}
and the azimuthal angle
\begin{equation}
\varphi_{\mathrm{a}} (\omega,\alpha) =
\tan^{-1} \left( \frac{\mathcal{V}^{\mathrm{ay}}(\omega,\alpha)}
{\mathcal{V}^{\mathrm{ax}}(\omega,\alpha)} \right),
\quad 0 \leq \varphi \leq 2\pi.
\label{eq:azimuthal-ang} 
\end{equation} 

\section{Results}
\label{sec:results}

\begin{table}[t]
\center
\begin{tabular}{ccccc}\\
\hline
\quad Atom \qquad & \multicolumn{3}{c}{Position (\AA)} \\
\cline{2-4}
\quad type \qquad & $x$ & $y$ & $z$  \\
\hline
H$_1$ & -0.615 & -1.774 &  0.731 \\
H$_2$ &  0.615 &  0.355 &  0.731 \\
C$_1$ & -0.615 & -1.772 & -0.491 \\
C$_2$ & -0.615 & -0.356 & -0.723 \\
C$_3$ &  0.615 &  0.357 & -0.490 \\
C$_4$ &  0.615 &  1.774 & -0.731 \\
\hline
\end{tabular}
\caption{
Atomic positions in the unit cell of the Up structure shown
in Fig. \ref{fig:up-struc}. 
}
\label{tab:up-unitcell}
\end{table}
\begin{table}[t]
\center
\begin{tabular}{cccc}\\
\hline
\quad Atom \qquad & \multicolumn{3}{c}{Position (\AA)} \\
\cline{2-4}
\quad type \qquad & $x$ & $y$ & $z$  \\
\hline
H$_1$ &  -0.615 &  -1.421 & \ 1.472 \\
C$_1$ &  -0.615 &  -1.733 & \ 0.396 \\
C$_2$ & \ 0.615 & \ 1.733 & \ 0.158 \\
C$_3$ & \ 0.615 & \ 0.422 &  -0.158 \\
C$_4$ &  -0.615 &  -0.373 &  -0.396 \\
H$_2$ &  -0.615 &  -0.685 &  -1.472 \\
\hline
\end{tabular}
\caption{
Atomic positions in the unit cell of the Alt structure
 shown in Fig. \ref{fig:alt-struc}. 
}
\label{tab:alt-unitcell}
\end{table}

We present the calculated results of
$\mathcal{V}_{\sigma^{\mathrm{b}}}(\omega,\alpha)$ and
$\mathcal{V}_{\mathrm{a}}(\omega,\alpha)$ for the Up  and
Alt structures, both noncentrosymmetric 2D carbon systems with
50\% hydrogenation, which are differently structurally arranged. We remind that
the Up structure has hydrogen atoms only on the upper side of the carbon sheet,
while the Alt structure has alternating hydrogen atoms on the upper and bottom
sides. We take the carbon lattice to be along the $xy$ plane for both
structures, and the carbon-hydrogen bonds are
perpendicular to  $xz$ plane for the Up structure
(Fig. \ref{fig:up-struc}),
 and off the normal for the Alt structure 
(Fig. \ref{fig:alt-struc}). The coordinates for the Up and Alt unit cells of the
structures are given in Tables \ref{tab:up-unitcell} and 
\ref{tab:alt-unitcell}, respectively.

We calculated the self-consistent ground state and the Kohn-Sham states within
density functional theory in the local density approximation (DFT-LDA), with a
planewave basis using the ABINIT code \cite{gonzeCPC09}. We used 
Hartwigsen-Goedecker-Hutter (HGH) relativistic separable dual-space Gaussian
pseudopotentials \cite{hartwigsenPRB98}, including the spin-orbit interaction
needed to calculate $\mu^{\mathrm{abcd}}(\omega,\alpha)$ from Eq.
\eqref{eq:mu}.
The convergence parameters for the calculations,  corresponding to the Up and
Alt structures are cutoff energies up to 65\,Ha, resulting in LDA
energy band gaps of 0.084\,eV and 0.718\,eV, respectively, 
and 14452 ${\mathbf k}$ points in the IBZ
where the energy
eigenvalues and matrix elements were calculated;
to
integrate $\mu^{\mathrm{abcd}}(\omega)$ and
$\xi^{\mathrm{a}\mathrm{b}}(\omega)$ the linearized analytic tetrahedron method
(LATM) has been used.\cite{nastosPRB07} 
We neglect the anomalous velocity term
$\hbar(\boldsymbol{\sigma}\times\nabla V)/4m^2c^2$, where $V$ is the crystal
potential, in $\hat{\mathbf v}$ of Eq.~\eqref{z.1}, as this term is known to
give small contribution to PSC.\cite{bhatPRL05} Therefore, $[\hat{\mathbf
v},\hat{\mathbf S}]=0$, and Eq.~\eqref{z.1} reduces to $\hat
K^{\mathrm{a}\mathrm{b}}=\hat v^\mathrm{a} \hat S^\mathrm{b}=\hat S^\mathrm{b}
\hat v^\mathrm{a}$. Finally, the prime in the sum of Eq.~\eqref{eq:mu} is
restricted to quasi-degenerated conduction bands $c$ and $c'$ that are closer
than 30 meV to each other, which  is both typical laser pulse energy width and the thermal
room-temperature energy level broadening.\cite{nastosPRB07}

\subsection{SVI: Spin velocity injection}
\label{sec:res-spin_velocity}

\begin{figure}[t]
\centering
\includegraphics[width=\tama]{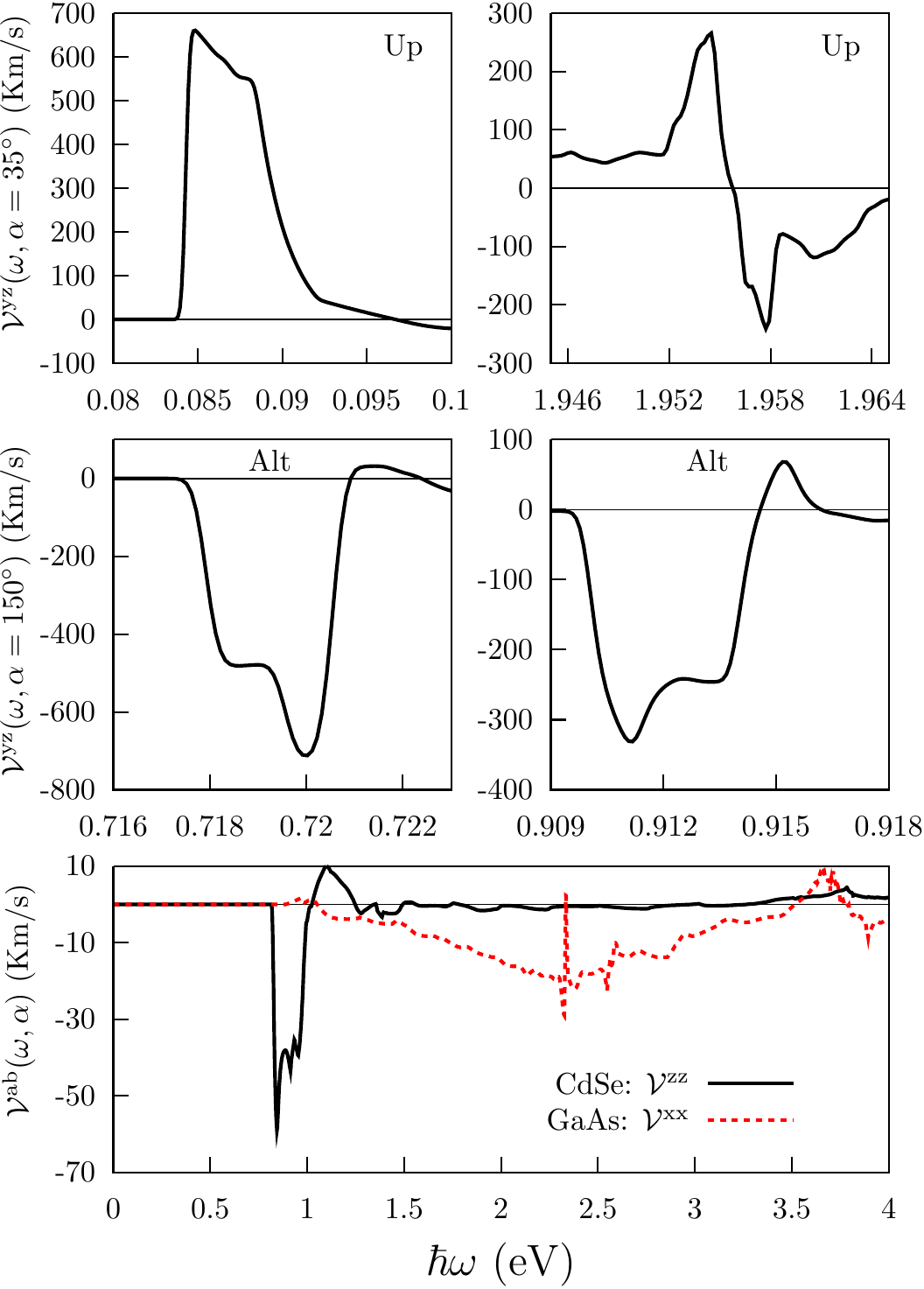}
\caption{(color online)  Spin velocity injection
  $\mathcal{V}^{\mathrm{ab}}(\omega,\alpha)$ {\it vs.} photon energy
  $\hbar\omega$, for the angles $\alpha$ that maximize the signal. The largest
  velocity are at the low energy regions of the spectra for the Alt and Up
  structures, becoming different from zero at the energy gap of each structure.
  In the high energy regions, the values of
  $\mathcal{V}^{\mathrm{ab}}(\omega,\alpha)$ are also very large compared to
  the 3D case of CdSe and GaAs, shown at the bottom panel.}
\label{fig:vab-str-comp}
\end{figure}

In Fig. \ref{fig:vab-str-comp}, we show $\mathcal{V}^{\mathrm{ab}}
(\omega,\alpha)$ {\it vs.} $\hbar\omega$ for the 
velocity and spin
directions
$\hat{\mathbf{a}}$ and $\hat{\mathbf{b}}$, and for the angle $\alpha$, at which
the signal is maximized, for the Up  and Alt structures, and for CdSe and GaAs
bulk systems, shown for comparison. As expected from
Eq.~\eqref{eq:mu}, 
${\cal V}^{\mathrm{a}\mathrm{b}}(\omega,\alpha)$ starts rising 
from zero right at the
corresponding energy gap of each system. For the 2D structures considered, the
spectrum contains two narrow energy regions with strong response, while for the
bulk systems, the spectra covers a rather wide energy range, but with a much
weaker response. For the Up structure, at $\mathrm{a}\mathrm{b}=yz$ and
$\alpha=35^\circ$ the response is maximized, which means that an incoming light
with its electric field polarized at 35$^\circ$ from the $x$ direction will
induce electrons to move along $y$ (parallel to the surface), with their spin
polarized along $z$ (perpendicular to the surface). Right at the energy onset, ${\cal
V}^{yz}(\omega,\alpha)=668$\,Km/s, remains almost constant for 65 meV, and then
decreases to zero. A second region with high velocity is above 1.946 eV with
two, opposite in sign, maxima of the speed: ${\cal
V}^{yz}(\omega,\alpha)=266.3$ Km/s at $\hbar\omega=1.954$\,eV, and ${\cal
V}^{\mathrm{a}\mathrm{b}}(\omega,\alpha)=-241.4$ Km/s at
$\hbar\omega=1.958$\,eV; a positive (negative) ${\cal
V}^{\mathrm{a}\mathrm{b}}(\omega,\alpha)$ means that the electrons move
parallel (antiparallel) to the electric field. Likewise, for the Alt structure,
we also find that $\mathrm{a}\mathrm{b}=yz$ and $\alpha=150^\circ$ maximizes
the response, where two extreme values of ${\cal V}^{\mathrm{y}\mathrm{z}}$ are
found, one at  $\hbar\omega=0.720$\,eV of ${\cal
V}^{\mathrm{y}\mathrm{z}}=-711.9$ Km/s, and the other at
$\hbar\omega=0.911$\,eV of ${\cal V}^{\mathrm{y}\mathrm{z}}=-330.6$ Km/s.
 
For the bulk structures, we calculate ${\cal V}^{\mathrm{a}\mathrm{b}}(\omega)$
from Eq.~\eqref{eq:vab-w} by simply using $\boldsymbol{\mu}_{\mathrm{max}}$.
For CdSe, we find that for $\hbar\omega=0.844$\,eV,
$\boldsymbol{\mu}_{\mathrm{max}}\to \mu^{zzzz}$, and ${\cal
V}^{\mathrm{z}\mathrm{z}}(\omega)=-59.0$ Km/s, and for GaAs at
$\hbar\omega=2.324$\,eV,
$\boldsymbol{\mu}_{\mathrm{max}}\to\mu^{\mathrm{aaaa}}$ and ${\cal
V}^{\mathrm{a}\mathrm{a}}(\omega,\alpha)=-28.7$\,Km/s, with $\mathrm{a}=x,y,z$.
For these bulk semiconductors, the $x$, $y$, and $z$ axis are taken along the
standard cubic unit cell directions, [100], [010], and [001], respectively. In
Table \ref{tab:vab-str-comp}, we compare ${\cal
V}^{\mathrm{a}\mathrm{b}}(\omega,\alpha)$ for the 2D structures considered and
bulk crystals. We stress that, as shown in the figure, the 2D structures
have maxima in ${\cal V}^{\mathrm{a}\mathrm{b}}(\omega;\alpha)$  higher than
for the bulk crystals by more than order of magnitude. 
In particular, the Alt structure demonstrates a 
${\cal V}^{\mathrm{a}\mathrm{b}}(\omega;\alpha)$ 
about 12 times larger than that
of CdSe and GaAs.

\begin{table}
\begin{tabular}{cccccc}
\hline
\multirow{2}{*}{Structure \quad} & 
System \quad & 
Pol. &
Energy & 
\multicolumn{2}{c}{$\mathcal{V}^{\mathrm{ab}}(\omega,\alpha)$}\\
\cline{5-6}
& type & Ang. & [eV] & $\mathrm{ab}$ \quad & [Km/s]\\
\hline
Up  & 2D   & 35    & 0.084  & $\mathrm{yz}$ &  660.5 \\
      &      &       & 1.954  & $\mathrm{yz}$ &  266.3 \\
      &      &       & 1.958  & $\mathrm{yz}$ & -241.4 \\
Alt & 2D   & 150   & 0.720  & $\mathrm{yz}$ & -711.9 \\
      &      &       & 0.911  & $\mathrm{yz}$ & -330.6 \\
CdSe  & bulk & --    & 0.844  & $\mathrm{zz}$ &  -59.0 \\
GaAs  & bulk & --    & 2.324  & $\mathrm{xx}$ &  -28.7 \\
\hline
\end{tabular}
\caption{Comparison of the reported maximum values of
$\mathcal{V}^{\mathrm{ab}}(\omega,\alpha)$ for the different structures and
their corresponding polarization angle $\alpha$ and energy $\hbar\omega$ . }
\label{tab:vab-str-comp}
\end{table}

\subsection{Fixing spin}
\label{sec:res-fixspin}

\begin{figure}[t]
\centering
\includegraphics[width=\tama]{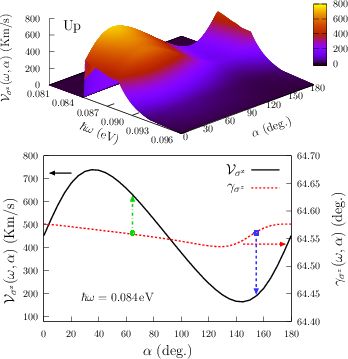}
\caption{(color online) For the Up structure, the top panel
shows ${\cal V}_{\sigma^\mathrm{z}}(\omega,\alpha)$ {\it vs.} $\hbar\omega$ and
$\alpha$, and the bottom panel shows
$\gamma_{\sigma^\mathrm{z}}(\omega,\alpha)$ (right scale, red short-dashed line), and ${\cal
V}_{\sigma^\mathrm{z}}(\omega,\alpha)$ (left scale, black solid line), {\it
vs.} $\alpha$, for $\hbar\omega=0.084$\,eV, i.e. along the ridge shown in the
3D plot. }
\label{fig:up-vsz-w1}
\end{figure}

\begin{figure}[t]
\centering
\includegraphics[width=\tama]{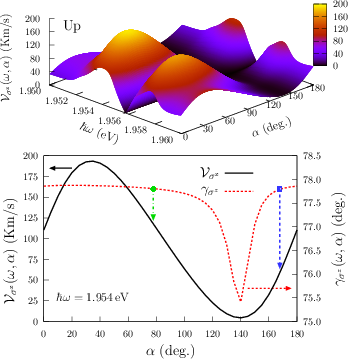}
\caption{(color online) For the Up structure, the top panel shows ${\cal
V}_{\sigma^\mathrm{z}} (\omega,\alpha)$ {\it vs.} $\hbar\omega$ and $\alpha$,
and the bottom panel shows $\gamma_{\sigma^\mathrm{z}}
(\omega,\alpha)$ 
(right scale, red short-dashed line), and ${\cal V}_{\sigma^\mathrm{z}}
(\omega,\alpha)$ (left scale, black solid line), {\it vs.} $\alpha$, for
$\hbar\omega=1.954$\,eV, i.e. along the highest ridge shown in the 3D plot. }
\label{fig:up-vsz-w2}
\end{figure}

In this subsection, we calculate ${\cal V}_{\sigma^\mathrm{z}}(\omega,\alpha)$,
Eq. \eqref{eq:vs-mag}, for the case with the spin fixed along $z$, i.e.,
directed perpendicularly to the surface of the Up and Alt structures. Also, we
calculate $\gamma_{\sigma^\mathrm{z}}(\omega,\alpha)$ from Eq. 
\eqref{eq:gamma-ang}, which determines the direction  of the injected 
electrons movement along the surface of  each structure. We mention that we
have also  analyzed the cases when the spin  is directed along $x$ or
$y$, finding similar qualitative results to those presented below.

\subsubsection{Up structure}\label{up:fs}

In the top panel of Fig. \ref{fig:up-vsz-w1}, we plot
$\mathcal{V}_{\sigma^{\mathrm{z}}} (\omega,\alpha)$ {\it vs.}
0.080\,eV$\leq\hbar\omega\leq$0.096\,eV (similar energy range for the Up
structure shown in the left panel of Fig.~\ref{fig:vab-str-comp}) and
$0^\circ\leq\alpha\leq 180^\circ$. We see a broad peak that reaches the maximum
of $\mathcal{V}_{\sigma^{\mathrm{z}}}(\omega,\alpha) = 739.7$\,Km/s at
$\alpha=35^{\circ}$ and $\hbar\omega= 0.084$\,eV.  The variation of
$\mathcal{V}_{\sigma^{\mathrm{z}}} (\omega,\alpha)$ as a function of $\alpha$,
which comes from the interplay of the $\boldsymbol{\mu}$ tensor components as
multiplied by the  trigonometric functions of Eq.~\eqref{eq:vab-aw}, gives a
sizable set of values between 739.7\,Km/s and 165.4\,Km/s, for
0.084\,eV$\leq\hbar\omega\leq$0.090\,eV. In the bottom panel, we show
$\mathcal{V}_{\sigma^{\mathrm{z}}} (\omega,\alpha)$ (left scale,
black solid line) {\it vs.} $\alpha$, at $\hbar\omega= 0.084$\,eV, thus following the ridge in the 3D
plot of the top panel. Also, we plot the corresponding velocity angle
$\gamma_{\sigma^\mathrm{z}}(\omega,\alpha)$ (right scale, red
short-dashed line), where it
is very interesting to see that $\gamma_{\sigma^z}(\omega,\alpha)$ is centered
at 64.55$^\circ$ with a rather small deviation of only $\pm 0.03^\circ$, for
the whole range of $\alpha$. This result means that for $\hbar\omega=0.084$ eV
and for all values of $\alpha$, the electrons, with the chosen spin pointing
along $z$, will move at the angle of $\gamma_{\sigma^\mathrm{z}}(\omega,\alpha)
\sim 64.5^{\circ}$ with respect to the $x$ direction, with the range of  high
speeds $\mathcal{V}_{\sigma^{\mathrm{z}}} (\omega,\alpha)$ shown in the figure.
Also, from Eq. \eqref{eq:gamma-par} we find that
$\gamma^\parallel_{\sigma^\mathrm{z}} (\omega,\alpha)=\alpha = 64.56^\circ$,
with $\mathcal{V}_{\sigma^{\mathrm{z}}} (\omega,\alpha) = 631.1$\,Km/s 
(as indicated by the green dot-dashed arrow), and that from Eq. \eqref{eq:gamma-perp},
$\gamma^\perp_{\sigma^\mathrm{z}}(\omega,\alpha)=\alpha-90^\circ=64.50^\circ$,
gives $\alpha=154.50^\circ$, with
$\mathcal{V}_{\sigma^{\mathrm{z}}}(\omega,\alpha) = 191.5$ Km/s 
(as indicated by the blue long-dashed 
arrow). 
Thus, at $\hbar\omega=0.084$ eV, an incident field, polarized at
$\alpha \sim 65.5^\circ$ or $\sim 154.5^\circ$, injects electrons with their
spin polarized along $z$, which move parallel or perpendicular to the incident
electric field,  with a speed of 631.14\,Km/s or 191.5\,Km/s, respectively.

Now, we analyze the results for the second energy range of the Up structure
shown in Fig.~\ref{fig:vab-str-comp}. In the top panel
of Fig. \ref{fig:up-vsz-w2}, we plot $\mathcal{V}_{\sigma^{\mathrm{z}}}
(\omega,\alpha)$ {\it vs.} 1.950\,eV$\leq\hbar\omega\leq$1.960\,eV and
$0^\circ\leq\alpha\leq 180^\circ$. We see two broad peaks that maximize at
$\alpha=35^{\circ}$ and $\hbar\omega= 1.954$\,eV, with a value of
$\mathcal{V}_{\sigma^{\mathrm{z}}}(\omega,\alpha) = 193.5$\,Km/s, and at
$\alpha=35^{\circ}$ and $\hbar\omega= 1.957$\,eV, with a value of
$\mathcal{V}_{\sigma^{\mathrm{z}}}(\omega,\alpha) = 170.6$\,Km/s. We only
analyze the highest  maximum in the bottom panel, where we  show
$\mathcal{V}_{\sigma^{\mathrm{z}}} (\omega,\alpha)$ (left scale,
black solid line) {\it vs.} $\alpha$, at $\hbar\omega= 1.954$\,eV, thus following the highest ridge
shown in the 3D plot of the top panel. Also, we plot the corresponding velocity
angle $\gamma_{\sigma^\mathrm{z}}(\omega,\alpha)$ (right scale, red short-dashed  line),
where in this case we see that the values of $\gamma_{\sigma^z}(\omega,\alpha)$
have more dispersion, as a function of $\alpha$, than for the lower energy
range shown in the bottom panel of Fig.~\ref{fig:up-vsz-w1}. However,
$\gamma_{\sigma^z}(\omega,\alpha)\sim 77.8^\circ$ is nearly constant from
$\alpha=0^\circ$ up to $\alpha\sim 85^\circ$. In this case, we find that
$\gamma^\parallel_{\sigma^\mathrm{z}}(\omega,\alpha)=\alpha=78.0^\circ$, with
$\mathcal{V}_{\sigma^{\mathrm{z}}}(\omega,\alpha) = 115.0$\,Km/s 
(as indicated by the green dot-dashed arrow), and that from Eq. \eqref{eq:gamma-perp},
$\gamma^\perp_{\sigma^\mathrm{z}}(\omega,\alpha)=\alpha-90^\circ=167.8^\circ$,
gives $\alpha=77.8^\circ$, with
$\mathcal{V}_{\sigma^{\mathrm{z}}}(\omega,\alpha) = 65.6$\,Km/s 
(as indicated by the blue long-dashed  arrow). 
Thus, through the correct choice of $\hbar\omega$ and $\alpha$ we could
inject electrons, in this case with their spin polarized along $z$, which move
parallel or perpendicular to the incident electric field, with sizable speeds.

\begin{figure}[tb]
\centering
\includegraphics[width=\tama]{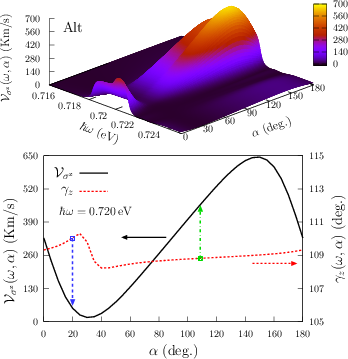}
\caption{(color online) For the Alt structure, the top panel
shows ${\cal V}_{\sigma^\mathrm{z}} (\omega,\alpha)$ {\it vs.} $\hbar\omega$
and $\alpha$, and the bottom panel shows $\gamma_{\sigma^\mathrm{z}}
(\omega,\alpha)$ (right scale, red short-dashed  line), and ${\cal V}_{\sigma^\mathrm{z}}
(\omega,\alpha)$ (left scale, black solid line), {\it vs.} $\alpha$, for
$\hbar\omega=0.720$\,eV, i.e. along the ridge shown in the 3D plot.}
\label{fig:alt-vsz}
\end{figure}

\subsubsection{Alt structure}

We proceed to analyze the Alt structure, just as we did with the Up structure,
but in this case, we only choose the lower energy range shown in the left
central panel of Fig.~\ref{fig:vab-str-comp}. In the top panel of Fig. 
\ref{fig:alt-vsz}, we plot $\mathcal{V}_{\sigma^{\mathrm{z}}} (\omega,\alpha)$
{\it vs.} 0.715\,eV$\leq\hbar\omega\leq$0.725\,eV and $0^\circ\leq\alpha\leq
180^\circ$. We see a broad peak that maximizes at $\alpha=150^{\circ}$ and
$\hbar\omega= 0.720$\,eV, with a value of
$\mathcal{V}_{\sigma^{\mathrm{z}}}(\omega,\alpha) = 644.9$\,Km/s. In the bottom
panel, we  show $\mathcal{V}_{\sigma^{\mathrm{z}}} (\omega,\alpha)$ (left scale, black solid line) {\it vs.}
$\alpha$, at $\hbar\omega= 0.720$\,eV, thus following
the highest ridge shown in the 3D plot of the top panel. Also, we plot the
corresponding velocity angle $\gamma_{\sigma^\mathrm{z}} (\omega,\alpha)$
(right scale, red short-dashed  line), where now we see that
$\gamma_{\sigma^z}(\omega,\alpha)$ is centered at $109.2^{\circ}$ having
variations of $\pm 1.0^{\circ}$ for $0^{\circ} \leq
\alpha \leq 180^{\circ}$. In this case, we find that
$\gamma^\parallel_{\sigma^\mathrm{z}} (\omega,\alpha) =
\alpha = 108.8^\circ$, with $\mathcal{V}_{\sigma^{\mathrm{z}}} (\omega,\alpha)
= 450.05$\,Km/s (as indicated by the  green dot-dashed arrow), 
and that from Eq. \eqref{eq:gamma-perp},
$\gamma^\perp_{\sigma^\mathrm{z}}(\omega,\alpha)=\alpha-90^\circ=110.0^\circ$,
gives $\alpha=20.0^\circ$, with $\mathcal{V}_{\sigma^{\mathrm{z}}}
(\omega,\alpha) = 60.84$\,Km/s 
(as indicated by the blue long-dashed  arrow). Thus, as for the Up structure,
we could inject electrons with a fixed spin, which moves parallel or
perpendicular to the incident electric field.

\subsection{Fixing the electron velocity} 
\label{sec:res-fixvel}

Here we calculated $\mathcal{V}_{\mathrm{a}}(\omega,\alpha)$ (Eq. 
\eqref{eq:vv-mag}) after fixing the electron velocity direction, $\hat{\mathbf{a}}$,
to the $x$ or $y$ direction along the surface of the Up and Alt structures,
and from Eqns. \eqref{eq:polar-ang} and \eqref{eq:azimuthal-ang}, 
we
determined the corresponding polar angle, $\theta_{\mathrm{a}}
(\omega,\alpha)$, and azimuthal angle, $\varphi_{\mathrm{a}}
(\omega,\alpha)$, of the resulting spin orientation.

\subsubsection{Up structure}

\begin{figure}[t]
\centering
\includegraphics[width=\tama]{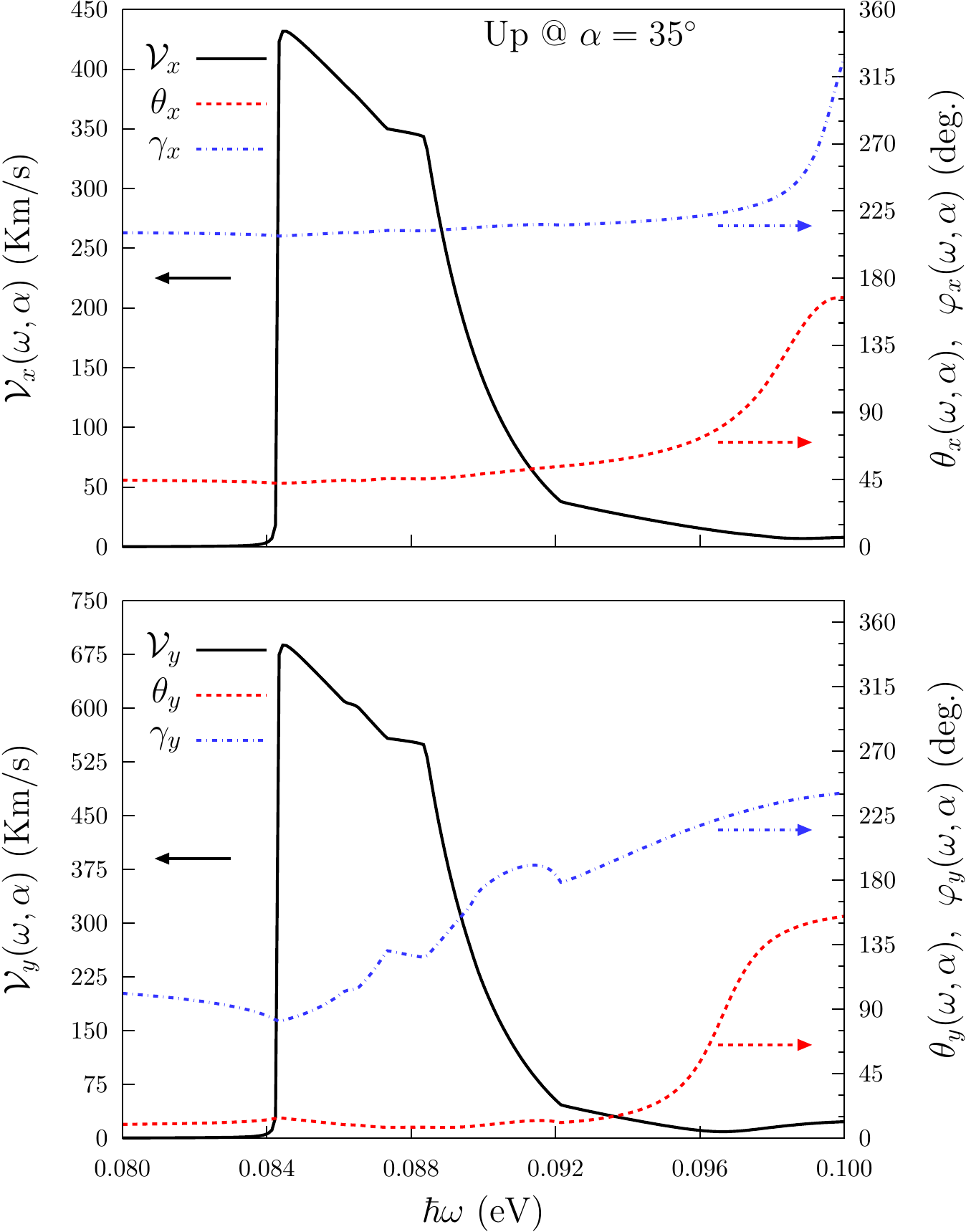}
\caption{(color online) For the Up structure we show the velocity
$\mathcal{V}_{\mathrm{a}} (\omega,\alpha)$ (left scale, black solid line), the polar
angle $\theta_{\mathrm{a}} (\omega,\alpha)$ (right scale, red dashed  line), and  the
azimuthal angle $\varphi_{\mathrm{a}} (\omega,\alpha)$, (right scale, blue dot-dashed 
line), {\it vs.} $\hbar\omega$, for $\alpha=35^\circ$, and $\mathrm{a}=x$ or
$\mathrm{a}=y$.}
\label{fig:up-vab-comp-rtp-1}
\end{figure}

For the Up structure, we find once again that  $\alpha=35^{\circ}$ maximizes
the response. In Fig. \ref{fig:up-vab-comp-rtp-1}, we plot
$\mathcal{V}_{\mathrm{a}} (\omega,\alpha)$ (left scale, black solid line),
$\theta_{\mathrm{a}} (\omega,\alpha)$ (right scale, red dashed  line), and
$\varphi_{\mathrm{a}} (\omega,\alpha)$, (right scale, blue dot-dashed  line), {\it vs.}
$\hbar\omega$, for $\mathrm{a}=x,y$. We see that for $\hbar\omega=0.084$\,eV,
the response has a maximum of $\mathcal{V}_{\mathrm{x}}
(\omega,\alpha)=431.7$\,Km/s at $\theta_{\mathrm{x}}(\omega,\alpha) =
42.5^{\circ}$, and $\varphi_{\mathrm{x}}(\omega,\alpha) = 208.3^{\circ}$, and
$\mathcal{V}_{\mathrm{y}} (\omega,\alpha)=687.9$\,Km/s at
$\theta_{\mathrm{y}}(\omega,\alpha) = 13.9^{\circ}$, and $\varphi_{\mathrm{y}}
(\omega,\alpha) = 82.1^{\circ}$. This means that the spin is directed upward
the third quadrant of the $xy$ plane when the electron moves along
$x$, and is almost parallel to the $xy$ plane in the first quadrant
when it moves along $y$. Also from this figure, we see that when the electron
moves along $x$, the spin direction is almost constant for all the energies
across the peak of the response, having $42.5^{\circ}<\theta_{\mathrm{x}}
(\omega,\alpha)<53.7^{\circ}$ and $208.3^{\circ}<\varphi_{\mathrm{x}}
(\omega,\alpha)<215.7^{\circ}$. When the electron moves along $y$, the spin
polar angle has again small variations, $11.3^{\circ}<
\theta_{\mathrm{y}}(\omega,\alpha)<13.9^{\circ}$, but the azimuthal angle
varies significantly, $82.1^{\circ}< \varphi_{\mathrm{y}}
(\omega,\alpha)<182.4^{\circ}$.

\begin{figure}[t]
\centering
\includegraphics[width=\tama]{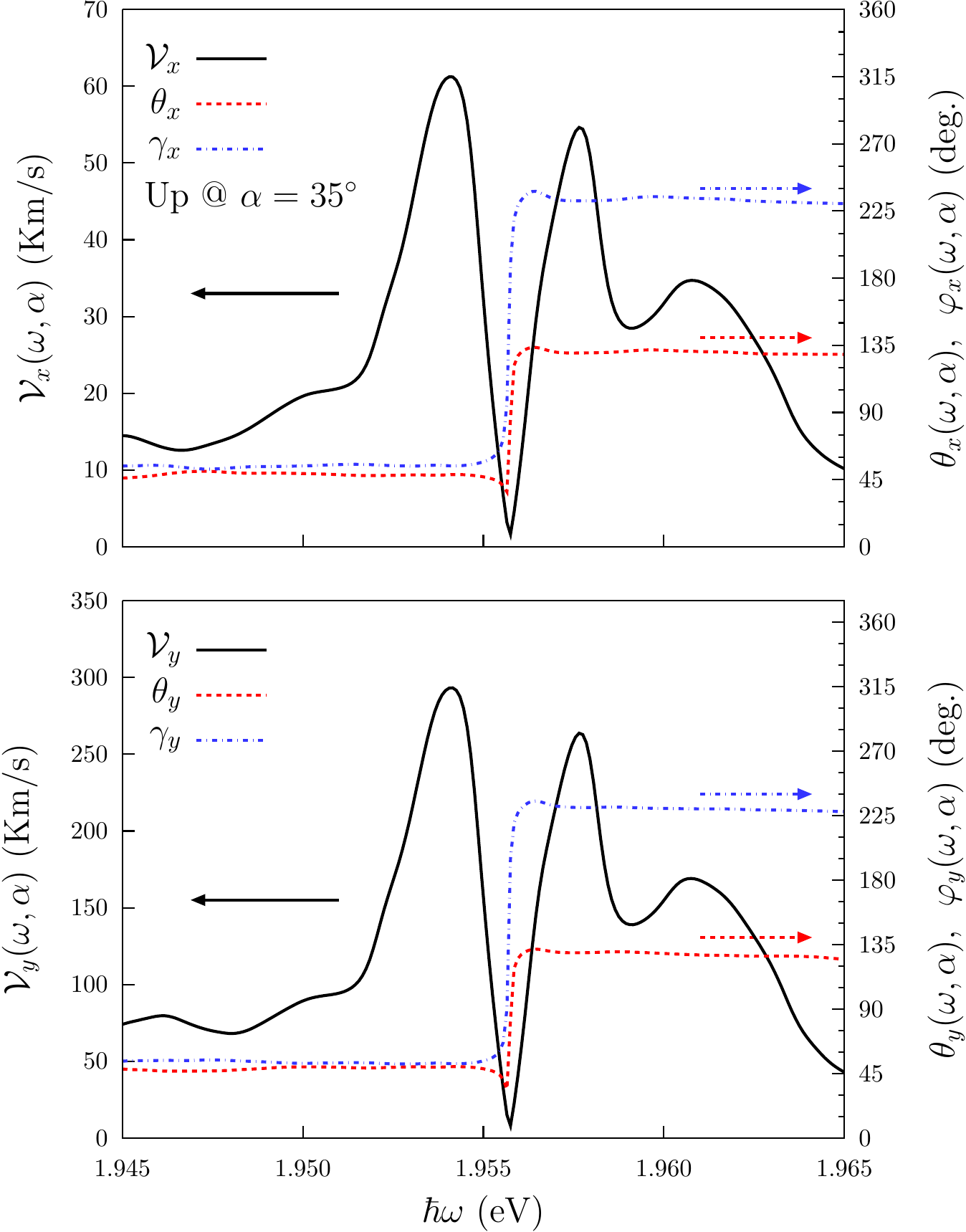}
\caption{(color online) For the Up structure we show the spin velocity
$\mathcal{V}_{\mathrm{a}} (\omega,\alpha)$ (left scale, black solid line), the polar
angle $\theta_{\mathrm{a}} (\omega,\alpha)$ (right scale, red dashed  line), and the
azimuthal angle $\varphi_{\mathrm{a}} (\omega,\alpha)$, (right scale, blue dot-dashed 
line), {\it vs.} $\hbar\omega$, for $\alpha=35^\circ$, and $\mathrm{a}=x$ or
$\mathrm{a}=y$. }
\label{fig:up-vx-vy-w2}
\end{figure}
In Fig. \ref{fig:up-vx-vy-w2}, we plot $\mathcal{V}_{\mathrm{a}}
(\omega,\alpha)$ {\it vs.} $\hbar\omega$, in the range where there two local maxima
with opposite sign at $\hbar\omega=1.954$\,eV and $\hbar\omega=1.957$\,eV
occur. The first is the largest of the two, with $\mathcal{V}_{\mathrm{x}}
(\omega,\alpha)=61.2$\,Km/s, $\theta_{\mathrm{x}}
(\omega,\alpha)=48.3^{\circ}$, and $\varphi_{\mathrm{x}}
(\omega,\alpha)=54.3^{\circ}$, for the electron moving along $x$; and
$\mathcal{V}_{\mathrm{y}} (\omega,\alpha)=293.2$\,Km/s, $\theta_{\mathrm{y}}
(\omega,\alpha)=49.8^{\circ}$, and $\varphi_{\mathrm{y}}
(\omega,\alpha)=51.9^{\circ}$ for the electron moving along $y$. For the peak
at $\hbar\omega=1.957$\,eV, we obtain $\theta_{\mathrm{x}} (\omega,\alpha) =
129.8^{\circ}$ and $\varphi_{\mathrm{x}} (\omega,\alpha) = 231.7^{\circ}$, with
$\mathcal{V}^{\mathrm{x}} (\omega,\alpha) = 54.6$\,Km/s and
$\theta_{\mathrm{y}}(\omega,\alpha) =129.3$; and
$\varphi_{\mathrm{y}}(\omega,\alpha) = 230.7$, with $\mathcal{V}^{\mathrm{y}}
(\omega,\alpha) = 263.7$\,Km/s. We remark that these angles are almost constant
for all the energy values across the peak of these two local maxima, for which
the spin is directed upward in the first quadrant of the $xy$ plane when
the electron moves along either $x$ or
$y$ directions.

\subsubsection{Alt structure}

\begin{figure}[tb]
\centering
\includegraphics[width=\tama]{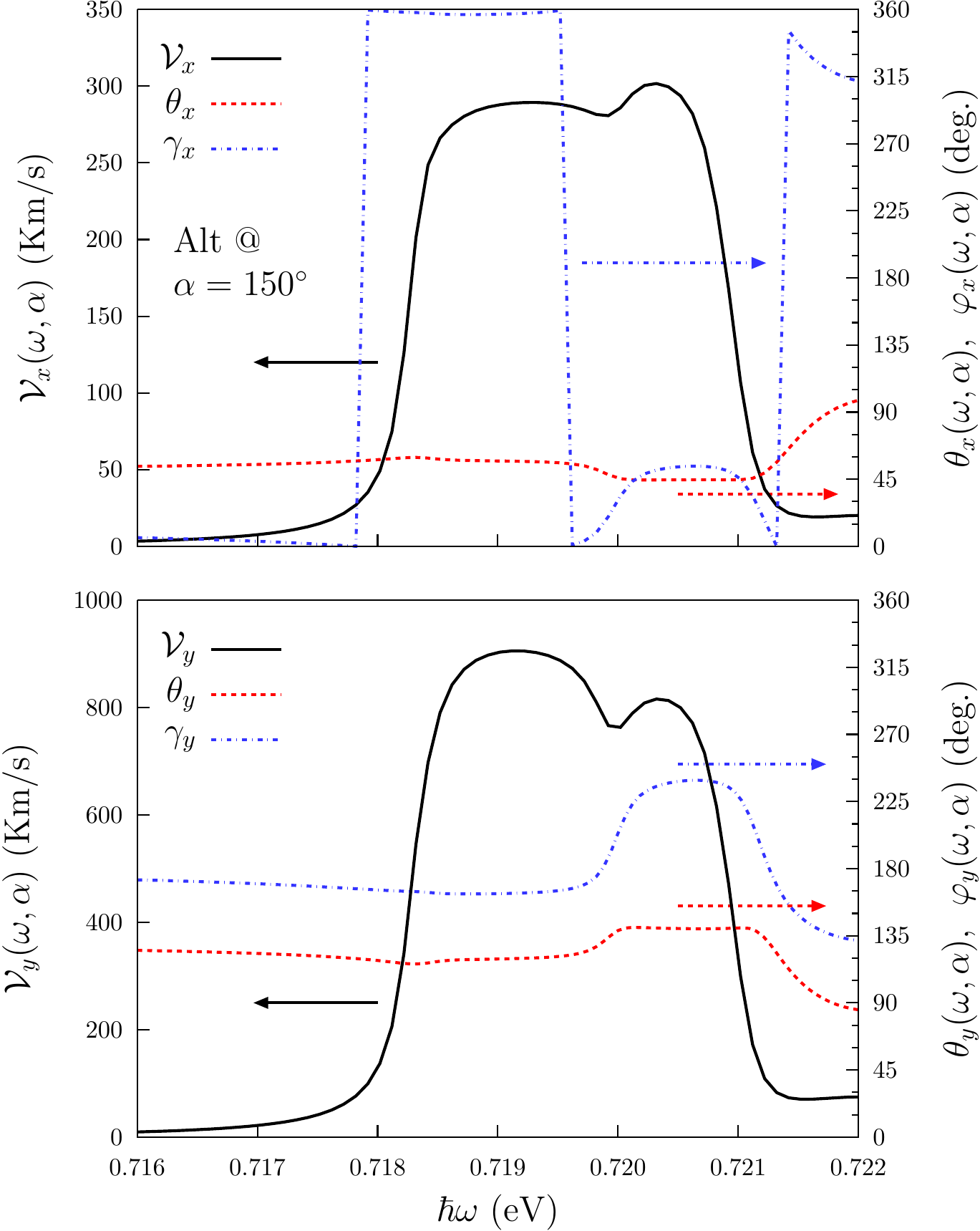}
\caption{(color online) For the Alt structure we show  the velocity
$\mathcal{V}_{\mathrm{a}} (\omega,\alpha)$ (left scale, black solid line), the polar
angle $\theta_{\mathrm{a}} (\omega,\alpha)$ (right scale, red dashed  line), and the
azimuthal angle $\varphi_{\mathrm{a}} (\omega,\alpha)$, (right scale, blue dot-dashed 
line), {\it vs.} $\hbar\omega$, for $\alpha=150^\circ$, and $\mathrm{a}=x$ or
$\mathrm{a}=y$. }
\label{fig:alt-vx-vy-w1}
\end{figure}

\begin{figure}[tb]
\centering
\includegraphics[width=\tama]{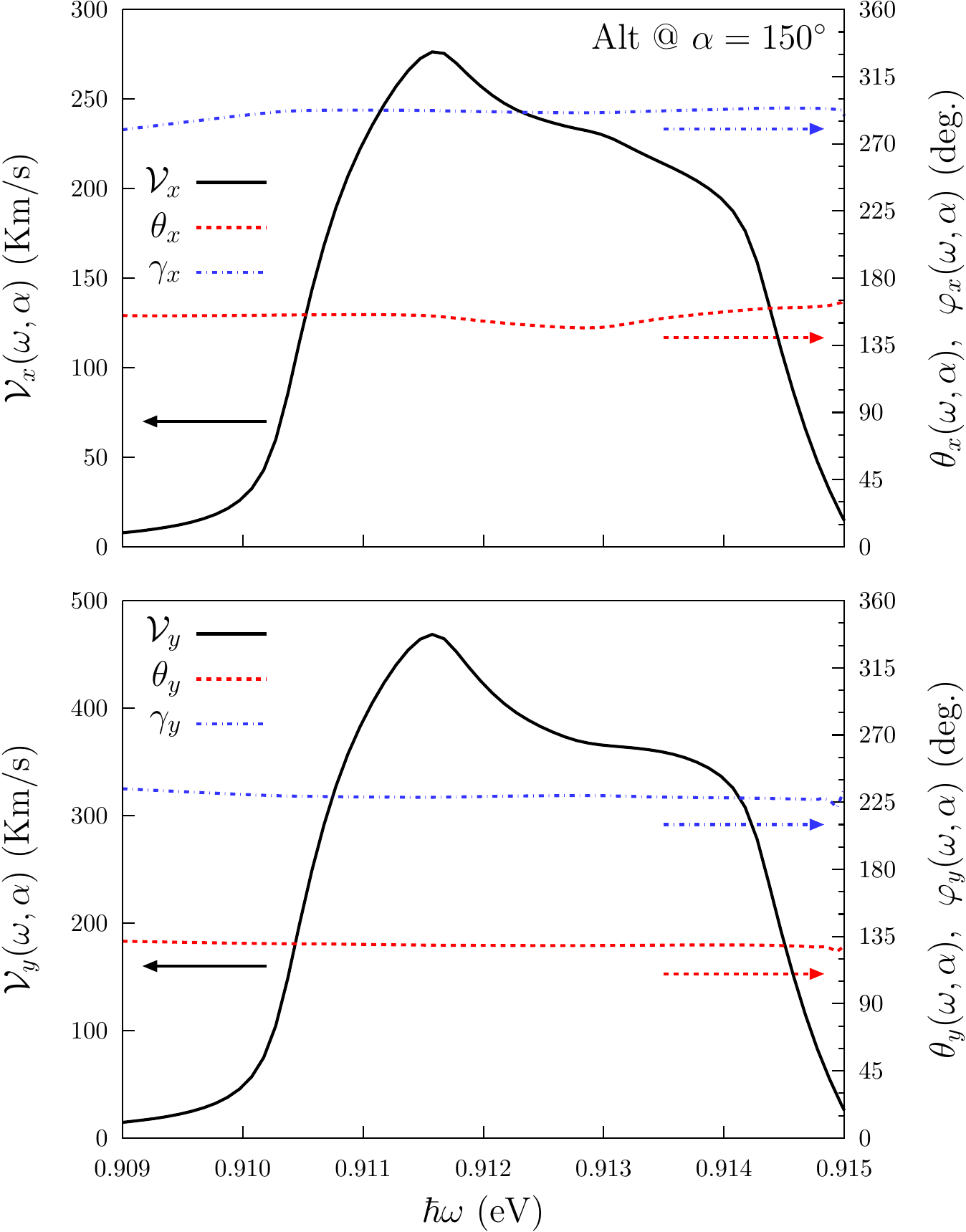}
\caption{(color online) For the Alt structure we show
$\mathcal{V}_{\mathrm{a}} (\omega,\alpha)$ (left scale, black solid line), the polar
angle $\theta_{\mathrm{a}} (\omega,\alpha)$ (right scale, red dashed  line), and the
azimuthal angle $\varphi_{\mathrm{a}} (\omega,\alpha)$, (right scale, blue dot-dashed 
line), {\it vs.} $\hbar\omega$, for $\alpha=150^\circ$, and $\mathrm{a}=x$ or
$\mathrm{a}=y$. }
\label{fig:alt-vx-vy-w2}
\end{figure}

In Figs. \ref{fig:alt-vx-vy-w1} and \ref{fig:alt-vx-vy-w2}, we plot
$\mathcal{V}_{\mathrm{a}} (\omega,\alpha)$ (left scale, black solid line),
$\theta_{\mathrm{a}} (\omega,\alpha)$ (right scale, red dashed line), and
$\varphi_{\mathrm{a}} (\omega,\alpha)$, (right scale, blue dot-dashed  line), {\it vs.}
$\hbar\omega$ in two different ranges, and  for $\mathrm{a}=x,y$. In this case,
$\alpha=150^\circ$  maximizes both $\mathcal{V}_{\mathrm{x}} (\omega,\alpha)$
and $\mathcal{V}_{\mathrm{y}} (\omega,\alpha)$, as a function of $\alpha$. In
Fig. \ref{fig:alt-vx-vy-w1}, the absolute maximum $\mathcal{V}_{\mathrm{x}}
(\omega,\alpha) = 301.7$\,Km/s is at $\hbar\omega=0.720$\,eV,
$\theta_{\mathrm{x}} (\omega,\alpha) = 44.5^{\circ}$ and
$\varphi_{\mathrm{x}}(\omega,\alpha) = 51.2^{\circ}$, and
$\mathcal{V}_{\mathrm{y}} (\omega,\alpha) = 905.6$\,Km/s at
$\theta_{\mathrm{y}} (\omega,\alpha) = 119.7^{\circ}$ and
$\varphi_{\mathrm{y}}(\omega,\alpha) = 163.4^{\circ}$. Thus, the spin is
directed upward the fourth quadrant of the $xy$ plane when the spin
velocity is directed along $x$, while it is directed downward the second
quadrant when the spin velocity is directed along $y$. 
Finally, in
Fig. \ref{fig:alt-vx-vy-w2}, the absolute maximum is at $\hbar\omega=0.911$\,eV
at $\mathcal{V}_{\mathrm{x}} (\omega,\alpha) = 276.3$\,Km/s,
$\theta_{\mathrm{x}} (\omega,\alpha) = 154.6^{\circ}$, and
$\varphi_{\mathrm{x}}(\omega,\alpha) = 292.3^{\circ}$, and
$\mathcal{V}_{\mathrm{y}} (\omega,\alpha) = 468.6$\,Km/s at
$\theta_{\mathrm{y}} (\omega,\alpha) = 129.2^{\circ}$, and
$\varphi_{\mathrm{y}}(\omega,\alpha) = 228.3^{\circ}$, implying that the spin
is directed downward the fourth quadrant of the $xy$ plane when the
spin velocity is directed along $x$, while is directed downward the third
quadrant when the spin velocity is directed along $y$.

\section{Conclusions}
\label{sec:conclusions}

We reported the results of an \emph{ab initio} calculations for the spin
velocity injection (SVI) due to the one-photon absorption of the linearly
polarized light in the Up and Alt 2D 50\% hydrogenated graphene structures.
Different possible arrangements of the of the spin injection have been
considered: we made the calculations for the cases when the spin is polarized
in $z$ direction or when the velocity is directed along $x$ or $y$. 
To the best  
of our knowledge, this effect  has not been previously reported in these  2D  
partially hydrogenated structures.  
We have shown that the  SVI demonstrates an
anisotropic behavior, which is very sensitive to the symmetry of the structures
of interest.
We have found that the Up structure shows the strongest response
for the spin directed along $z$, resulting in the velocity
$\mathcal{V}_{\sigma^{\mathrm{z}}} (\omega,\alpha) = 668.0$\,Km/s for the
incoming photon energy of 0.084\,eV. Also, the Alt structure has the strongest
response when the spin moves along $y$ direction, resulting in
$\mathcal{V}^{\mathrm{y}} (\omega,\alpha) = 905.6$\,Km/s for the incoming
photon energy of 0.720\,eV. The speed values obtained here are of the same
order of magnitude as those of Ref.~\onlinecite{najmaiePRB03} in unbiased
semiconductor quantum well structures, while they are of order of magnitude
higher compared to 3D bulk materials. 
Considering the fact that the spin relaxation time in pure and
doped graphene ranges from nanoseconds to milliseconds,
\cite{wojtaszekPRB13,ertlerPRB09}, and in view of the
high spin velocity transport that we obtained for  both
structures, this time is sufficiently long enough to have the SVI
effect observed experimentally. 
Therefore, the Up and the Alt graphene structures
considered here are  excellent
candidates for the development of spintronics devices that require pure spin
current (PSC).

\section{Acknowledgment}
This work has been supported by \emph{Consejo Nacional de Ciencia y
Tecnolog\'ia} (CONACyT), M\'exico, Grant No. 153930.
R.Z.P. thanks CONACyT for scholarship support.
A.I.S thanks to Centro de Investigaciones en Optica (CIO) for the               
hospitality during his sabbatical research leave.

\end{document}